# $\mathcal{H}_\infty$ Performance Analysis for Almost Periodic Piecewise Linear Systems with Application to Roll-to-Roll Manufacturing Control

Christopher Martin, Edward Kim, Enrique Velasquez, Wei Li, and Dongmei Chen, *Member, IEEE*

*Abstract*—An almost periodic piecewise linear system (APPLS) is a type of piecewise linear system where the system cyclically switches between different modes, each with an uncertain but bounded dwell-time. Process regulation, especially disturbance rejection, is critical to the performance of these advanced systems. However, a method to guarantee disturbance rejection has not been developed. The objective of this study is to develop an $\mathcal{H}_\infty$ performance analysis method for APPLSs, building on which an algorithm to synthesize practical $\mathcal{H}_\infty$ controllers is proposed. As an application, the developed methods are demonstrated with an advanced manufacturing system – roll-to-roll (R2R) dry transfer of two-dimensional materials and printed flexible electronics. Experimental results show that the proposed method enables a less conservative and much better performing $\mathcal{H}_\infty$ controller compared with a baseline $\mathcal{H}_\infty$ controller that does not account for the uncertain system switching structure.

## I. INTRODUCTION

Many engineering systems can be modeled as a piecewise linear system, where the system behavior is defined by different linear equations in different conditions. When the system behavior repeats periodically over time, these systems are called periodic piecewise linear systems (PPLSs) [1]–[4]. In many cases, the dwell times, which are the amounts of time that the system spends in each system mode, of PPLSs are uncertain but bounded [5], [6]. Such systems are called almost periodic piecewise linear systems (APPLSs), as the dwell-time uncertainty introduces slight aperiodicity [5], [7]. Figure 1 illustrates the switching structures of a PPLS and an APPLS, both with $S$ number of modes. In the PPLS case, the switching time between system modes is deterministic. For APPLSs, switching occurs within bounded time intervals, represented by the segments labeled $i||i+1$ for switching from mode $i$ to mode $i+1$, where $i = 1, 2, 3, …$, and $S$. Note that while there is bounded uncertainty in the switching times and thus the dwell-time of each system mode, the switching sequence in an APPLS is known.

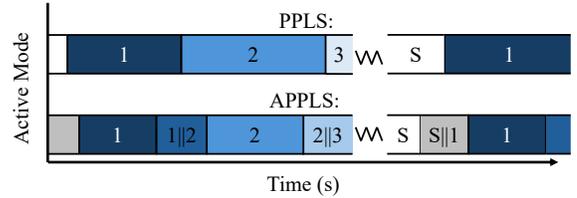

**Fig. 1.** The PPLS vs. APPLS switching structure. There are S system modes. Each mode has a distinct linear dynamic model.

An example of an APPLS is the newly developed roll-to-roll (R2R) manufacturing process for dry transfer of two-dimensional (2D) materials and printed electronics [8]–[10]. Figure 2 shows a schematic of the process [10]–[12]. The functional material is first laminated between the donor and the receiver substrates. The laminate is then unwound, passed through a set of guiding rollers, and peeled apart so that the functional material delaminates from the donor substrate while adhering to the receiver substrate. During the peeling process, the functional materials to be transferred often exhibit a pattern. For example, in [13] an array of $MoS_2$ strips were peeled from silicon, and in [10], [12] a series of graphene samples were peeled from a copper growth substrate. The abrupt transitions between different sections in these patterns can be modeled with different linear dynamics using a switched system structure. Since the peeling pattern is known in advance, the system follows a fixed switching sequence. However, in R2R systems, position error (also known as register error) is a common issue [14], making the switching times, and equivalently the dwell-times, uncertain. Fortunately, there will be tolerance limits on register errors in industrial R2R applications, i.e., the dwell-time uncertainty will be bounded [6], [15].

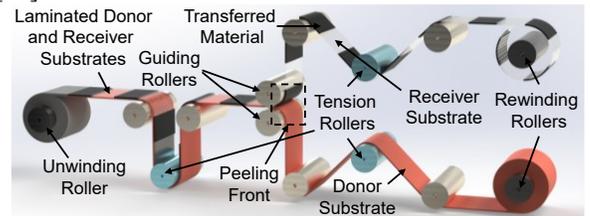

**Fig. 2.** The R2R dry transfer process

This work is based upon work supported primarily by the National Science Foundation under Cooperative Agreement No. EEC-1160494 and CMMI-2041470. Any opinions, findings and conclusions or recommendations expressed in this material are those of the author(s) and do not necessarily reflect the views of the National Science Foundation. (Corresponding author: Christopher Martin)

Christopher Martin, Edward Kim, Enrique Velasquez, Wei Li, and Dongmei Chen are with The Walker Department of Mechanical Engineering, The University of Texas at Austin, Austin, TX, USA, (cbmartin129@utexas.edu, edwardkim8775@utexas.edu, enrique.velasquez@utexas.edu, weiwli@austin.utexas.edu, dmchen@me.utexas.edu)

Previous research on APPLSs has been focused on stability and stabilization methods based on linear matrix inequality (LMI) techniques [5]–[7], [16], [17]. Due to high-precision requirements for advanced APPLS systems, disturbance rejection becomes a critical control objective to maintain the system performance [15], [18]. To design such a controller, it is first necessary to analyze the $\mathcal{H}_\infty$ performance, which represents the worst-case disturbance rejection capability of the closed-loop system [19]–[21]. However, no $\mathcal{H}_\infty$ performance analyses for APPLSs have been found in the literature.

The need for a dedicated $\mathcal{H}_\infty$ analysis method for APPLSs arises from their unique switching structure. Existing $\mathcal{H}_\infty$ analysis methods for switched periodic systems do not account for dwell-time uncertainty [16], [22]. The method developed in this study represents the first rigorous $\mathcal{H}_\infty$ performance analysis framework for cyclically switched linear systems that accounts for bounded dwell-time uncertainty. In addition to the dwell-time uncertainty, the analysis is also extended to handle norm-bounded modeling uncertainties [23], [24], making it generalizable to uncertain APPLSs [6] where each system mode is allowed to have additive modeling uncertainty. Furthermore, a controller synthesis method is developed to optimize the $\mathcal{H}_\infty$ performance of uncertain APPLSs. The developed method is experimentally validated on an R2R dry transfer testbed for 2D materials.

**Notation:** The Euclidean norm for vectors and the corresponding induced norm for matrices are denoted by $\|\cdot\|$. Scalars are represented using regular (non-bold) font (e.g., $a$, $B$, $\Gamma$), while vectors are written in bold lowercase font (e.g., $\boldsymbol{v}$, $\boldsymbol{w}$), and matrices are expressed in bold uppercase font (e.g., $\boldsymbol{A}$, $\boldsymbol{B}$). The identity matrix is denoted by $\boldsymbol{I}$.

## II. PROBLEM FORMULATION

### A. The uncertain APPLS modeling structure

The uncertain APPLS under consideration is an APPLS with additive modelling uncertainty in each mode. The almost periodic switching structure of such systems is cyclic with bounded dwell-time uncertainties. Explicitly, when $t \in [T_P^* + t_{l,i-1}, T_P^* + t_{l,i})$, an uncertain APPLS can be represented as

$$\dot{\boldsymbol{x}}(t) = (\boldsymbol{A}_i + \boldsymbol{F}_i \boldsymbol{\Delta}(t) \boldsymbol{G}_i) \boldsymbol{x}(t) + \boldsymbol{B}_i \boldsymbol{u}(t) + \boldsymbol{B}_{w_i} \boldsymbol{w}(t)$$
$$\boldsymbol{z}(t) = \boldsymbol{C}_i \boldsymbol{x}(t) + \boldsymbol{D}_i \boldsymbol{u}(t) \quad (1)$$

where $\boldsymbol{x}(t) \in \mathbb{R}^n$, $\boldsymbol{u}(t) \in \mathbb{R}^{n_u}$, $\boldsymbol{w}(t) \in \mathbb{R}^{n_w}$, and $\boldsymbol{z}(t) \in \mathbb{R}^{n_z}$, are the state, control input, disturbance input, and system output, respectively. $\boldsymbol{F}_i \in \mathbb{R}^{n \times n}$ and $\boldsymbol{G}_i \in \mathbb{R}^{n \times n}$ are the uncertainty weights in mode $i$, and $\boldsymbol{\Delta}(t) \in \mathbb{R}^{n \times n}$ is uncertain and time-varying such that $\|\boldsymbol{\Delta}(t)\| \le 1$. Thus, $\boldsymbol{F}_i$ and $\boldsymbol{G}_i$, $i = 1, 2, \ldots,$ and $S$, parameterize the mode-dependent additive modeling uncertainty. Additionally, $T_P^*$ is the length of a period, $lT_P^* + t_{l,i}$ is the switching time from mode $i$ to mode $i + 1$ in the $l^{th}$ period, and $lT_P^* + t_{l,S}$ is the switching time from mode $S$ to mode 1 in the $l^{th}$ period. These period-dependent switching times are unknown but bounded, so that $t_{l,i} \in [\underline{t_i}, \overline{t_i})$. Without loss of generality, let each switching pattern begin with the time segment where mode 1 must be active. Then, $0 \le \underline{t_1} \le t_{l,1} \le \overline{t_1} \le \underline{t_2} \le t_{l,2} \le \overline{t_2} \le \cdots \le \underline{t_S} \le t_{l,S} \le \overline{t_S} = T_P^*$. Thus, $\underline{t_i}$ and $\overline{t_i}$, $i = 1, 2, \ldots,$ and $S$, parameterize the dwell-time uncertainty. Define $T_i = \underline{t_i} - \overline{t_{i-1}}$ and $T_{i,i+1} = \overline{t_i} - \underline{t_i}$ as the durations of the time segments where mode $i$ must be active and where the switch from mode $i$ to $i + 1$ occurs, respectively. For a cyclically switched system without dwell-time uncertainty (a PPLS), $T_i$ is the dwell-time in mode $i$ and $T_{i,i+1} = 0$. This uncertain APPLS modeling structure is illustrated in Fig. 3, along with the APPLS and PPLS modeling structures for comparison purposes. The shaded region is where the uncertain linear dynamics can exist over time, while the solid line is an example trajectory.

### B. Switching control structure

To match the almost periodic switching structure, the control structure considered in this study is chosen to be switched, full-state feedback [6] as below,

$$\boldsymbol{u}(t) = \begin{cases} \boldsymbol{K}_i \boldsymbol{x}(t), & t \in [lT_P^* + \overline{t_{i-1}}, lT_P^* + \underline{t_i}) \\ \boldsymbol{K}_{i,i+1} \boldsymbol{x}(t), & t \in [lT_P^* + \underline{t_i}, lT_P^* + \overline{t_i}) \end{cases} \quad (2)$$

where the feedback gains, $\boldsymbol{K}_i$, $\boldsymbol{K}_{i,i+1}$, $i = 1, 2, \ldots,$ and $S$, are pre-calculated offline. The block diagram of the uncertain APPLS modeling structure in Eq. (1) and switched controller structure in Eq. (2) is summarized in Fig. 4.

The control objective here is to mitigate the impact of $\boldsymbol{w}$ on $\boldsymbol{z}$. To achieve this goal, the next section introduces a novel $\mathcal{H}_\infty$ performance analysis method for uncertain APPLSs. This result is then used to develop a novel $\mathcal{H}_\infty$ controller with the structure given in Eq. (2).

## III. CONTROL DESIGN

### A. Novel $\mathcal{H}_\infty$ performance analysis for uncertain APPLSs

Theorem 1 establishes that the uncertain APPLS described in Eq. (1), when controlled using the switching structure defined in Eq. (2), achieves a guaranteed weighted $\mathcal{H}_\infty$ performance, as defined in Definition 1 below, provided it satisfies the specified set of matrix inequalities

**Definition 1:** The uncertain APPLS given in Eq. (1) is said to have a weighted $\mathcal{H}_\infty$ performance $\gamma$ if, for $\lambda > 0, \gamma > 0$, and $b > 0$, it is exponentially stable when $\boldsymbol{w} = \boldsymbol{0}$, and the following holds for zero initial conditions and any non-zero $\boldsymbol{w}$ with finite energy:

$$\frac{1}{b} \int_0^\infty e^{-\lambda \tau} \boldsymbol{z}^T(\tau) \boldsymbol{z}(\tau) d\tau \le \gamma^2 \int_0^\infty \boldsymbol{w}^T(\tau) \boldsymbol{w}(\tau) d\tau \quad (3)$$

**Theorem 1:** Let there be $\alpha_i \ge 0$, $\alpha_{i,i+1} \ge 0$, $\alpha_{i+1,i+1} \ge 0$, $\boldsymbol{Q}_i > 0$, $\boldsymbol{Q}_{i,i+1} > 0$, $\boldsymbol{Y}_i$, $\boldsymbol{Y}_{i,i+1}$, $\lambda^* > 0$, and given constants $\lambda_i$, $\lambda_{i,i+1}, \mu_i \ge 1$, and $\mu_{i,i+1} \ge 1$ such that, for $i = 1, 2, \ldots,$ and $S$,

$$\begin{bmatrix} \boldsymbol{\Phi}_i & * & * \\ \boldsymbol{C}_i \boldsymbol{Q}_i + \boldsymbol{D}_i \boldsymbol{Y}_i & -\gamma^2 \boldsymbol{I} & 0 * \\ \boldsymbol{G}_i \boldsymbol{Q}_i & 0 & -\alpha_i \boldsymbol{I} \end{bmatrix} \le 0 \quad (4.A)$$

$$\begin{bmatrix} \boldsymbol{\Phi}_{i,i+1} & * & * \\ \boldsymbol{C}_i \boldsymbol{Q}_{i,i+1} + \boldsymbol{D}_i \boldsymbol{Y}_{i,i+1} & -\gamma^2 \boldsymbol{I} & * \\ \boldsymbol{G}_i \boldsymbol{Q}_{i,i+1} & 0 & -\alpha_{i,i+1} \boldsymbol{I} \end{bmatrix} \le 0 \quad (4.B)$$

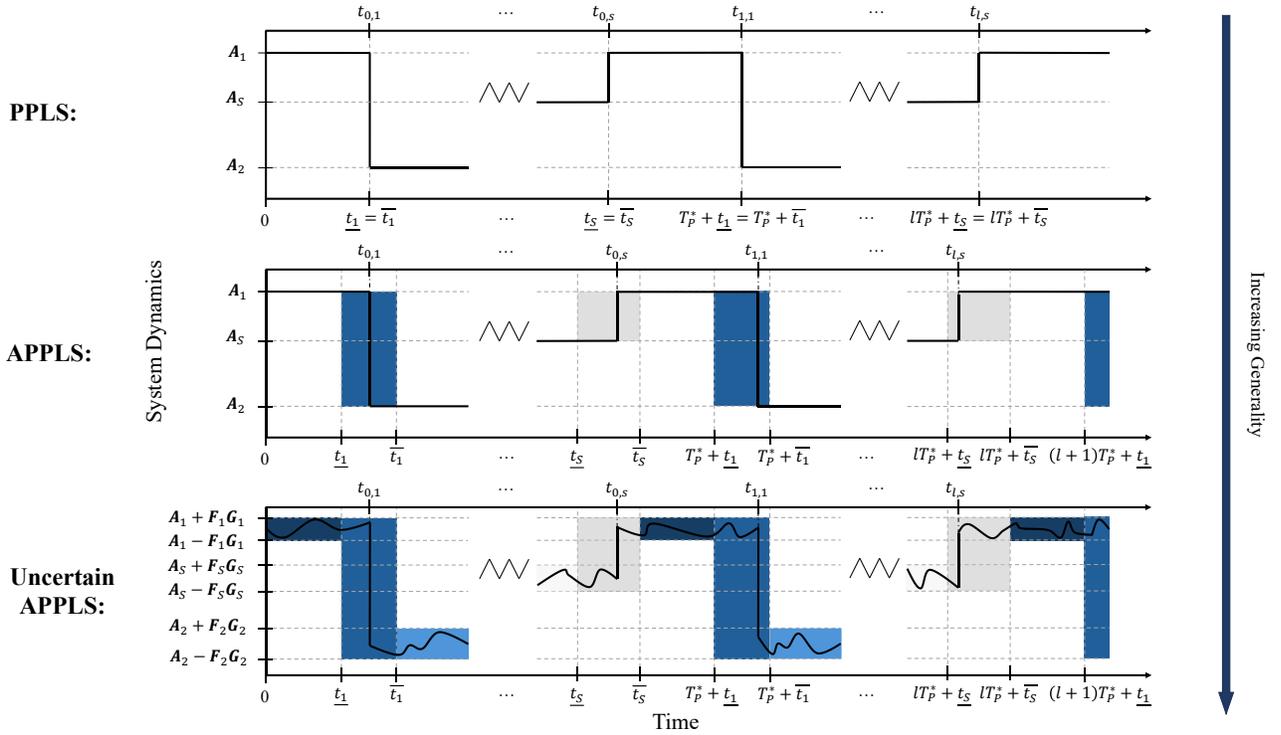

**Fig. 3.** The PPLS, APPLS, and uncertain APPLS modeling structures.

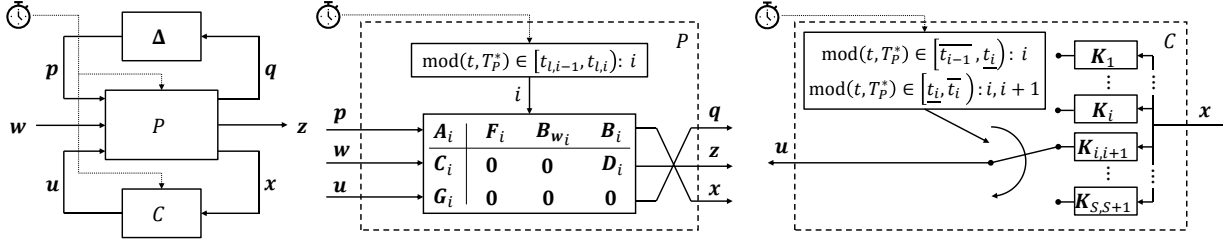

**Fig. 4.** Block diagram of uncertain APPLS modeling and control structure.

$$\begin{bmatrix} \boldsymbol{\Phi}_{i+1,i+1} & * & * \\ \boldsymbol{C}_{i+1}\boldsymbol{Q}_{i,i+1} + \boldsymbol{D}_{i+1}\boldsymbol{Y}_{i,i+1} & -\gamma^2 \boldsymbol{I} & * \\ \boldsymbol{G}_{i+1}\boldsymbol{Q}_{i,i+1} & 0 & -\alpha_{i+1,i+1}\boldsymbol{I} \end{bmatrix} \leq 0 \quad (4.C)$$

$$\boldsymbol{Q}_{S,S+1} \leq \mu_1 \boldsymbol{Q}_1 \quad (5.A)$$
$$\boldsymbol{Q}_{i-1,i} \leq \mu_i \boldsymbol{Q}_i \quad (5.B)$$
$$\boldsymbol{Q}_i \leq \mu_{i,i+1} \boldsymbol{Q}_{i,i+1} \quad (5.C)$$
$$\sum_{i=1}^{S} \Lambda_i - \mathrm{M}_i \geq 2\lambda^* T_P^* \quad (6)$$

are satisfied, where $\boldsymbol{\Phi}_i = sym(\boldsymbol{A}_i\boldsymbol{Q}_i + \boldsymbol{B}_i\boldsymbol{Y}_i) + \lambda_i \boldsymbol{Q}_i + \boldsymbol{B}_{w_i}\boldsymbol{B}_{w_i}^T + \alpha_i \boldsymbol{F}_i \boldsymbol{F}_i^T$, $\boldsymbol{\Phi}_{i,i+1} = sym(\boldsymbol{A}_i\boldsymbol{Q}_{i,i+1} + \boldsymbol{B}_i\boldsymbol{Y}_{i,i+1}) + \lambda_{i,i+1} \boldsymbol{Q}_{i,i+1} + \boldsymbol{B}_{w_i}\boldsymbol{B}_{w_i}^T + \alpha_{i,i+1} \boldsymbol{F}_i \boldsymbol{F}_i^T$, $\boldsymbol{\Phi}_{i+1,i+1} = sym(\boldsymbol{A}_{i+1}\boldsymbol{Q}_{i,i+1} + \boldsymbol{B}_{i+1}\boldsymbol{Y}_{i,i+1}) + \lambda_{i,i+1} \boldsymbol{Q}_{i,i+1} + \boldsymbol{B}_{w_{i+1}}\boldsymbol{B}_{w_{i+1}}^T + \alpha_{i+1,i+1} \boldsymbol{F}_{i+1} \boldsymbol{F}_{i+1}^T$, $\Lambda_i = \lambda_i T_i + \lambda_{i,i+1} T_{i,i+1}$, $\mathrm{M}_i = \ln \mu_i + \ln \mu_{i,i+1}$, and $sym(\boldsymbol{\Omega}) = \boldsymbol{\Omega} + \boldsymbol{\Omega}^T$. Then, for the uncertain APPLS in Eq. (1) with zero initial conditions, controlled using the switching structure in Eq. (2), with control gains $\boldsymbol{K}_i = \boldsymbol{Y}_i \boldsymbol{Q}_i^{-1}$ and $\boldsymbol{K}_{i,i+1} = \boldsymbol{Y}_{i,i+1} \boldsymbol{Q}_{i,i+1}^{-1}$, the weighted $\mathcal{H}_\infty$ performance given in Definition 1 can be guaranteed, where $\lambda = \lambda_{max}$, $b = \frac{\lambda_{max}}{2\lambda^*} e^{\left((\lambda^* + \frac{\lambda_{max}}{2} - \lambda_{min})2T_P^*\right)}$, $\lambda_{min} = \min\{\lambda_1, \lambda_{1,2}, \lambda_2, \ldots, \lambda_S\}$, and $\lambda_{max} = \max\{\lambda_1, \lambda_{1,2}, \lambda_2, \ldots, \lambda_S\}$. Note that for Eq. (6) to hold, $\lambda_{max} > 0$.

**Proof:** See Appendix. In the proof, the following mode-dependent Lyapunov function (MDLF) is used [6], [7].

$$V(t) = \begin{cases} x(t)^T \boldsymbol{P}_i x(t), \ t \in \left[lT_P^* + \overline{t_{i-1}}, lT_P^* + \underline{t_i}\right) \\ x(t)^T \boldsymbol{P}_{i,i+1} x(t), \ t \in \left[lT_P^* + \underline{t_i}, lT_P^* + \overline{t_i}\right) \end{cases} \quad (7)$$

where $\boldsymbol{P}_i = \boldsymbol{P}_i^T > 0$ and $\boldsymbol{P}_{i,i+1} = \boldsymbol{P}_{i,i+1}^T > 0$. In Eqs. (4) and (5), $\boldsymbol{Q}_i = \boldsymbol{P}_i^{-1}$ and $\boldsymbol{Q}_{i,i+1} = \boldsymbol{P}_{i,i+1}^{-1}$.

**Remark 1:** The $\mathcal{H}_\infty$ performance bound in Definition 1 is weighted by a decaying exponential to facilitate the use of the discontinuous MDLF in Eq. (7). This practice is common in the literature for switched systems [16], [25], as these MDLF discontinuities allow the $\mathcal{H}_\infty$ performance analysis method to be applied to a broader class of systems.

**Remark 2:** Theorem 1 can be readily extended to other MDLF matrix structures, such as continuous time-varying [7], or discontinuous time-varying [16].

**Remark 3:** The LMIs in Theorem 1 align with the switching structure of APPLSs. Equation (4.A) bounds the system output during the time segments when mode $i$ must be active, and Eqs. (4.B) and (4.C) bound the system output during the uncertain time segments when the switch from mode $i$ to $i + 1$ must occur. Additionally, Eq. (5) bounds the growth of the system when switching between these time segments.

*B. Weighted $\mathcal{H}_\infty$ controller synthesis*

Utilizing Theorem 1, a practical $\mathcal{H}_\infty$ controller synthesis algorithm for the uncertain APPLS given in Eq. (1) with the control structure in Eq. (2) is presented below in Algorithm 1. The algorithm has two phases; the first iterates until a desired stability margin is achieved, and the second iterates to optimize the weighted $\mathcal{H}_\infty$ performance. In this way stability can be guaranteed, while disturbance rejection can be maximized. For the stability phase, the following stabilizing LMI result will be used [6]. If there is $\alpha_i \geq 0$, $\alpha_{i,i+1} \geq 0$, $\alpha_{i+1,i+1} \geq 0$, $Q_i \geq 0$, $Q_{i,i+1} \geq 0$, $Y_i$, $Y_{i,i+1}$, and given constants $\lambda_i, \lambda_{i,i+1}, \mu_i \geq 1$, and $\mu_{i,i+1} \geq 1$ such that,

$$\begin{bmatrix} sym(A_i Q_i + B_i Y_i) + \alpha_i F_i F_i^T + \lambda_i Q_i & * \\ G_i Q_i & -\alpha_i I \end{bmatrix} \quad (8.A)$$

$$\begin{bmatrix} sym(A_i Q_{i,i+1} + B_i Y_{i,i+1}) \\ +\alpha_{i,i+1} F_i F_i^T + \lambda_{i,i+1} Q_{i,i+1} & * \\ G_i Q_{i,i+1} & -\alpha_{i,i+1} I \end{bmatrix} \quad (8.B)$$

$$\begin{bmatrix} sym(A_{i+1} Q_{i,i+1} + B_{i+1} Y_{i,i+1}) \\ +\alpha_{i+1,i+1} F_{i+1} F_{i+1}^T + \lambda_{i,i+1} Q_{i,i+1} & * \\ G_{i+1} Q_{i,i+1} & -\alpha_{i+1,i+1} I \end{bmatrix} \quad (8.C)$$

and Eqs. (5) and (6) hold, then, assuming $w = 0$, it can be readily proven that the uncertain APPLS in Eq. (1) is $\lambda^*$-exponentially stabilized by the control law in Eq. (2) with $K_i = Y_i Q_i^{-1}$, and $K_{i,i+1} = Y_{i,i+1} Q_{i,i+1}^{-1}$. More details can be found in [6]. Additionally, the following practical constraint on the norm of the controller gain is introduced. If

$$Q > I, \begin{bmatrix} -c^2 I & Y^T \\ Y & -I \end{bmatrix} < 0 \quad (9)$$

then $\|K\| < c$, where $K = YQ^{-1}$ [26]. In Algorithm 1, Eq. (9) will be applied to the pairs $Q_i, Y_i$ and $Q_{i,i+1}, Y_{i,i+1}$ to ensure that the norms of $K_i$ and $K_{i,i+1}$ are bounded.

| Algorithm 1: $\mathcal{H}_\infty$ Control Design (HCD) |
|---|
| **Step 0:** Choose $\epsilon > 0, \lambda_{lim} \geq 0, \mu_i \geq 1, \mu_{i,i+1} \geq 1, c > 0$, and $M_\lambda > 0$ |
| **Phase 1: Guarantee Convergence** |
| **Step 1.1:** Set $\chi^0 = +\infty$ and $k = 1$. **If** Eqs. (8.A) and (9) hold for some $Q_i \geq 0$ and $Y_i$ when $\lambda_i = 0$: $\lambda_i^0 = 0$. **Else:** $\lambda_i^0 = -M_\lambda$. **If** Eqs. (8.B), (8.C), and (9) hold for some $Q_{i,i+1} \geq 0$ and $Y_{i,i+1}$ when $\lambda_{i,i+1} = 0$: $\lambda_{i,i+1}^0 = 0$. **Else:** $\lambda_{i,i+1}^0 = -M_\lambda$. |
| **Step 1.2:** With $\lambda_i^{k-1}, \lambda_{i,i+1}^{k-1}$; **find** $Q_i^k, Q_{i,i+1}^k, Y_i^k, Y_{i,i+1}^k$, such that Eqs. (5), (8), and (9), hold. |
| **Step 1.3:** With $Q_i^k, Q_{i,i+1}^k, Y_i^k, Y_{i,i+1}^k$; find $\lambda_i^k, \lambda_{i,i+1}^k$ that **minimize** $\chi^k = -\sum_{i=1}^{S}(\lambda_i^k T_i + \lambda_{i,i+1}^k T_{i,i+1})$ subject to Eqs. (5) and (8). |
| **Step 1.4: If** $\lambda^* \geq \lambda_{lim}$ or $\|\chi^k - \chi^{k-1}\| < \epsilon$: go to Phase 2. **Else:** $k+=1$, return to Step 1.2. |
| **Phase 2: Optimize Weighted $\mathcal{H}_\infty$ Performance** |
| **Step 2.1:** $\lambda_i^0 = \lambda_i, \lambda_{i,i+1}^0 = \lambda_{i,i+1}$ from end of Phase 1. $\chi^0 = +\infty$. |
| **Step 2.2:** With $\lambda_i^{k-1}, \lambda_{i,i+1}^{k-1}$; **minimize** $\gamma^k$ over $Q_i^k, Q_{i,i+1}^k, Y_i^k, Y_{i,i+1}^k$, such that Eqs. (4), (5) and (9) hold. |
| **Step 2.3:** With $\gamma^k, Q_i^k, Q_{i,i+1}^k, Y_i^k, Y_{i,i+1}^k$; find $\lambda_i^k, \lambda_{i,i+1}^k$ that **minimize** $\chi^k$ subject to Eqs. (4) and (5). |
| **Step 2.4: If** $\|\chi^k - \chi^{k-1}\| < \epsilon$: STOP. **Else:** $k+=1$, return to Step 2.2. |
| **Stability Check:** If $\lambda^* > 0$, where $\lambda^*$ is from the end of Phase 2, then the feedback gains $K_i = Y_i Q_i^{-1}$ and $K_{i,i+1} = Y_{i,i+1} Q_{i,i+1}^{-1}$ achieve a weighted $\mathcal{H}_\infty$ performance $\gamma$ for the system defined by Eqs. (1) and (2). Also, for $w = 0$, the same system is $\lambda^*$-exponentially stable. |

**Remark 4:** The novel $\mathcal{H}_\infty$ control framework presented in Theorem 1 and Algorithm 1 is designed for uncertain APPLSs, and it can also readily be applied to APPLSs, since uncertain APPLSs are a generalization of APPLSs.

**Remark 5:** Algorithm 1 can be used to achieve optimal performance without significant online computation since the uncertain APPLS structure assumes known mode-dependent dynamics and unpredictable disturbance inputs, making adaptive $\mathcal{H}_\infty$ control approaches [27], [28] unnecessary and methods that learn disturbance characteristics online [29] inapplicable. Thus, the $\mathcal{H}_\infty$ controller synthesis can achieve optimal performance while maintaining a practical structure that is compatible with industrial applications that have limited online computational resources.

## IV. CASE STUDY: R2R DRY TRANSFER

The uncertain APPLS analysis and control framework developed in Section III is applied to the R2R dry transfer of patterned materials, addressing the system switching dynamics and mode-dependent uncertainties. Experimental results demonstrate the effectiveness of the novel $\mathcal{H}_\infty$ controller in improving peeling angle control and overall system performance.

*A. The R2R dry transfer patterned peeling dynamic model*

This subsection summarizes the R2R dry transfer patterned peeling dynamic model [6], [9], [30]. Fig. 5 depicts the peeling front of the R2R dry transfer system, and Table I lists the physical parameters. In the figure, different web colors represent different web material sections. Each material section is represented as a different dynamic mode in the system model, because certain material parameters will change depending on which section of the web is being peeled. These pattern section-dependent variables are listed at the end of the table.

The web velocity derivatives of the R2R dry transfer system are given as follows,

$$\dot{v}_j(t) = -\frac{R_j^2}{J_j} t_j(t) + \frac{R_j}{J_j} u_j(t) - \frac{f_j}{R_j} v_j(t), \quad j = 2,3 \quad (10)$$

The web tension dynamics can be determined as a function of the unstretched length derivatives as

$$\dot{t}_j = \frac{\partial t_j}{\partial l_1}(t) \cdot \dot{l}_1 + \frac{\partial t_j}{\partial l_2}(t) \cdot \dot{l}_2 + \frac{\partial t_j}{\partial l_3}(t) \cdot \dot{l}_3, \quad j = 2,3 \quad (11)$$

where

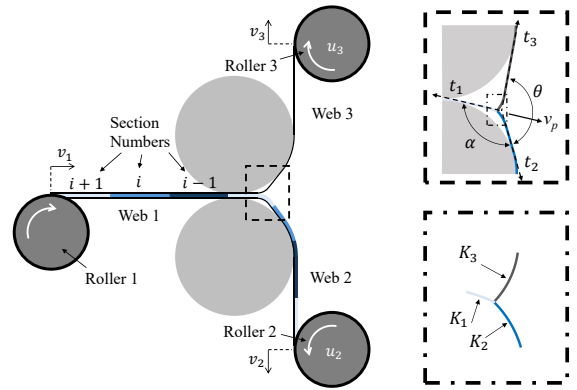

**Fig. 5.** R2R dry transfer of a patterned material.

## TABLE I
### R2R PEELING PARAMETERS

| Variable | Description (Units) |
|---|---|
| $t_j, j = 1, 2, 3$ | Tension in web $j$ (N) |
| $v_j, j = 1, 2, 3$ | Velocity of web $j$ (m/s) |
| $l_j, j = 1, 2, 3$ | Unstretched length web $j$ (m) |
| $\theta, \alpha$ | Peeling angles (radians) |
| $\varepsilon_j, j = 1, 2, 3$ | Strain in web $j$ (m/m) |
| $u_j, j = 2, 3$ | Motor torque inputs (N-m) |
| $v_p$ | Peeling front velocity (m/s) |
| $K_j, j = 1, 2, 3$ | Bending curvature of web $j$ (1/m) |
| $E_j, j = 1, 2, 3$ | Elastic modulus of web $j$ (N/m$^2$) |
| $R_j, j = 1, 2, 3$ | Radius of roller $j$ (m) |
| $J_j, j = 1, 2, 3$ | Moment of inertia roller $j$ (kg-m$^2$) |
| $f_j, j = 1, 2, 3$ | Friction coefficient of roller $j$ (m/s) |
| Pattern Section-Dependent Parameters | |
| $I_{j,i}, j = 1, 2, 3$ | Moment of Inertia of web $j$ section $i$ (m$^4$) |
| $\Gamma_i$ | Adhesion Energy of section $i$ (J/m$^2$) |
| $b_i$ | Width of the contact surface of section $i$ (m) |

$$\dot{l}_1(t) = \frac{v_1(t) - v_p(t)}{1 + \varepsilon_1(t)} \quad (12.A)$$

$$\dot{l}_j(t) = \frac{v_p(t)}{1+\varepsilon_1(t)} - \frac{v_j(t)}{1+\varepsilon_j(t)}, \quad j = 2, 3 \quad (12.B)$$

The unstretched lengths are proportional to the amount of mass in a web and the sensitivities $\frac{\partial t_j}{\partial l_k}$ can be found numerically [15], [18]. Finally, assuming constant peeling and neglecting higher-order terms, the energy balance at the peeling front can be summarized as

$$t_2 + t_3 - t_1 = \tau \quad (13)$$

where

$$\tau = b_i \Gamma_i - \frac{1}{2}\left[E_2 I_{2,i}\left(K_1^2 - K_2^2\right) + E_3 I_{3,i}\left(K_1^2 - K_3^2\right)\right] (14)$$

### B. Representing R2R peeling as an uncertain APPLS

Next, the peeling angles, which are illustrated in Fig. 5, can be found as a function of the three web tensions.

$$\theta = \mathrm{acos}\frac{t_1^2 - t_2^2 - t_3^2}{2t_2 t_3} \quad (15.A)$$

$$\alpha = \mathrm{acos}\frac{t_3^2 - t_1^2 - t_2^2}{2t_1 t_2} \quad (15.B)$$

The peeling angles have been shown to have a direct impact on the transfer quality of the R2R dry transfer process [10], [12]. In addition, controlling these angles is challenging, as even small changes in web bending energy and adhesion can lead to significant errors. Thus, the primary control goal in this study is to regulate the peeling angles at a set point [8]–[10], [13], [31].

Using Eqs. (10)-(15), the nonlinear state dynamics can be compactly represented as follows,

$$\dot{x}(t) = f_i(x(t), w(t), u(t)) \quad (16)$$

where $x = [v_2, v_3, \theta, \alpha]^T$ are the system states, $w = [\tau, v_p]^T$ are the disturbance inputs, and $u = [u_2, u_3]^T$ are the control inputs. When mode $i$ is active, the nonlinear system dynamics in Eq. (16) can be relaxed into the following linear differential inclusion (LDI) [6], [15], [18]:

$$\delta\dot{x}(t) \in Co(\mathcal{A}_i)\delta x(t) + B_i \delta u(t) + B_{w_i}\delta w(t) \quad (17)$$

where $\delta x = x - \tilde{x}, \delta u = u - \tilde{u}$, and $\delta w = w - \tilde{w}$; and $\tilde{x}, \tilde{u}$, and $\tilde{w}$ are reference signals. Additionally, $B_i = \frac{\partial f_i}{\partial u}$ is constant, $B_{w_i} = \frac{\partial f_i}{\partial w}$ is treated as constant since $\frac{\partial f_i}{\partial w}$ does not vary much with respect to $w$, $Co(\cdot)$ is the convex hull operator, and $\mathcal{A}_i$ is the LDI set defined as follows,

$$\mathcal{A}_i = \left\{\frac{\partial f_i}{\partial x}\bigg|_{\tilde{x},\tilde{u}} \bigg| v_p \in \left[\underline{v_p}, \overline{v_p}\right], \tau \in \left[\underline{\tau_i}, \overline{\tau_i}\right]\right\} \quad (18)$$

where $\underline{v_p}, \overline{v_p}$ and $\underline{\tau_i}, \overline{\tau_i}$ are the lower and upper bounds of $v_p$ and $\tau$, respectively. Note that, since $I_{j,i}, j = 2, 3, b_i$, and $\Gamma_i$ are mode-dependent, the bounds on $\tau$ are also mode-dependent, while the bounds on $v_p$ are not. This LDI can be represented as a linear system with norm-bounded uncertainty [15].

$$\delta\dot{x}(t) = (A_i + F_i \Delta(t) G_i)\delta x(t) + B_i \delta u(t) + B_{w_i}\delta w(t) \quad (19)$$

Thus, each material mode can be naturally modeled as a linear system with additive, norm-bounded uncertainty.

To determine the switching signal, it is assumed that the peeling follows a predictable pattern with $S$ peeling modes, such that mode $i + 1$ is peeled immediately after mode $i$, as in Fig. 5. In addition, it is assumed that this pattern is cyclic, so that mode 1 is peeled immediately after mode $S$. Let $v_1$ be constant and define $\underline{q_i}$ and $\overline{q_i}$ as the closest and farthest distance, relative to the start of the pattern, that mode $i$ can switch to mode $i + 1$, respectively. Under these assumptions, the R2R patterned peeling system can be represented as an uncertain APPLS, where

$$T_P^* = \overline{q_S}/v_1 \quad (20.A)$$

$$\underline{t_i} = \underline{q_i}/v_1, \overline{t_i} = \overline{q_i}/v_1 \quad (20.B)$$

and the linear modes with modeling uncertainty are given in Eq. (19).

Figure 6 summarizes the process of representing R2R patterned peeling as an uncertain APPLS. Figure 6(a) shows the system with two sections, or modes. Figure 6(b) plots the bounds on the uncertain variables, $\tau$ and $v_p$, over time. According to Eqs. (18) and (19), these bounds can be used to represent each dynamic mode as a linear system with additive modelling uncertainty. In addition, the switching times on these bounds, derived using Eq. (20), also define the uncertain switching times of the system. Finally, Fig. 6(c) illustrates the uncertain APPLS modeling structure for this patterned peeling system.

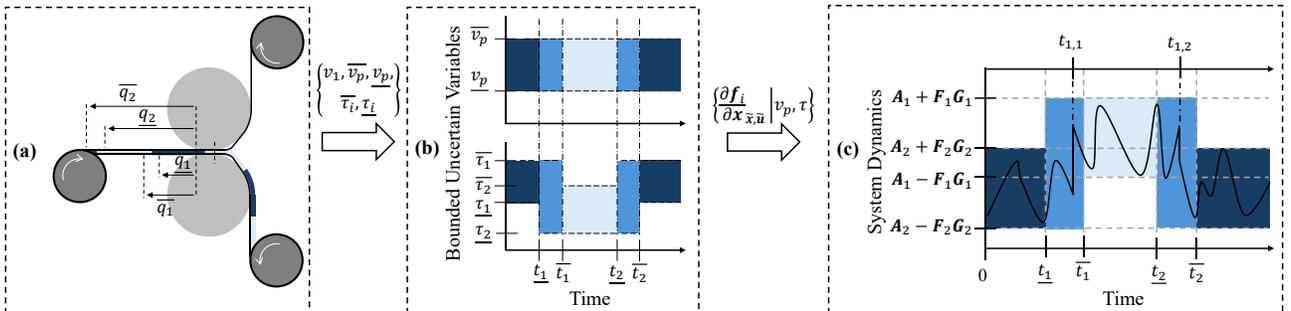

**Fig. 6.** Representing R2R dry transfer of patterned materials as an uncertain APPLS.

*C. Baseline controller design and controller tuning*

The proposed $\mathcal{H}_\infty$ controller for uncertain APPLSs, developed in Section III, is applied to regulate R2R patterned peeling on an experimental testbed, representing the first experimental validation of a switching control method for R2R peeling. In this study, the control objective is to regulate the system states defined in Eq. (16) around an optimal setpoint, prioritizing the critical peeling angles to ensure their precise control. The performance of the proposed controller is compared against a baseline time-invariant, non-switching $\mathcal{H}_\infty$-optimal full-state-feedback controller. This controller is the only $\mathcal{H}_\infty$ controller that has been applied to R2R peeling in the literature [15]. The block diagram of this baseline controller is shown in Fig. 7.

This baseline controller is synthesized and implemented without utilizing the a-priori knowledge of the peeling pattern, instead relying on a linear time invariant (LTI) model of the system with additive norm-bounded uncertainty. The nominal linear dynamics of the baseline system are defined as the average of the system matrices across all modes, $(A, B_w, B) = \text{mean}_i(A_i, B_{w_i}, B_i)$. Additionally, the additive modelling uncertainty, parameterized by $F$ and $G$, is constructed under the assumption that $\tau$ can vary between its maximum and minimum bounds across all peeling modes. As a result, the baseline controller is overly conservative, while the proposed controller properly incorporates the a-priori knowledge of the peeling pattern.

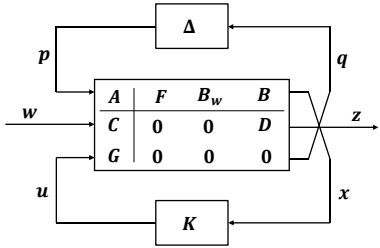

**Fig. 7.** Block diagram for the baseline controller.

To ensure fairness of comparison, the $c$ parameter from Eq. (9) was adjusted independently so that, for both controllers, $b\gamma^2 = 1$, where $b$ is given in Theorem 1. Thus, both controllers guarantee the same $\mathcal{H}_\infty$ performance. Additionally, the $C_i$ and $D_i$ matrices, which weigh the cost of state error and control effort, as described in Eq. (1), are identical for both controllers. Specifically, $C_i = \begin{bmatrix} Q_x \\ 0 \end{bmatrix}$ and $D_i = \begin{bmatrix} 0 \\ R_u \end{bmatrix}$ where $Q_x$ and $R_u$ are diagonal matrices tuned using Bryson's rule [32], such that $Q_{x_{ii}} \sim \tilde{x}_i^{-2}$ and $R_{u_{ii}} \sim \tilde{u}_i^{-2}$. Additional weight was put on the cells in $Q_x$ that correspond to the peeling angle errors. Also, $\epsilon = 10^{-2}$, $\mu_i = \mu_{i,i+1} = 1$, and $M_\lambda = 10^3$ for both controllers. Since the control goal is to reject the disturbances caused by $\tau$ and $v_p$, $\lambda_{lim} = 0$ so that $\gamma$ will be as small as possible while still ensuring stability.

*D. Experimental setup*

The experimental setup of the R2R testbed is shown in Fig. 8. The roller velocities and web tensions are measured using encoders on the two rewinding rolls and loadcells on each of the three web sections, as illustrated in Fig. 8(a). The control hardware uses a National Instrument CompactRIO to collect these measurements [15]. The tension measurements are used to calculate $\alpha$, $\theta$, and $\tau$ online according to Eqs. (13)-(15). The controller produces voltage outputs to actuate the brushless motors that drive the two rewinding rollers.

Figure 8(b) shows the peeling pattern. Two different peeling modes were created using two types of scotch tape peeled from a polyethylene terephthalate (PET) web. The two types of tape have significantly different adhesion properties. Figure 8(b) also depicts the relationship between switching parameters ($q_i$, $\overline{q_i}$, $T_i$, and $T_{i,i+1}$) and the material mode switch locations. For each period of each experimental run, the switch from mode 1 to mode 2 is randomly chosen to be between $\underline{q_1}$ and $\overline{q_1}$, simulating the dwell-time uncertainty characteristic of advanced R2R processes. An analogous approach is used to choose the switch locations from mode 2 to mode 1. The physical parameters of the setup are given in Table II [15].

Experiments were conducted under three different operating cases. The angle setpoints, unwinding velocities, and peeling patterns of each case are given in Table III. Six experiments were run for each case. The conditions of this experiment are relevant to any R2R peeling process, as the key dynamic factors—such as fluctuating and unpredictable adhesion energy, bending energy, and peeling front velocity—closely resemble those in peeling advanced 2D materials. For example, Eqs. (13) and (14), which were used to design the controllers, were originally developed to model the R2R dry transfer of graphene [9]. Thus, the result can be scaled as necessary for industrial R2R peeling processes.

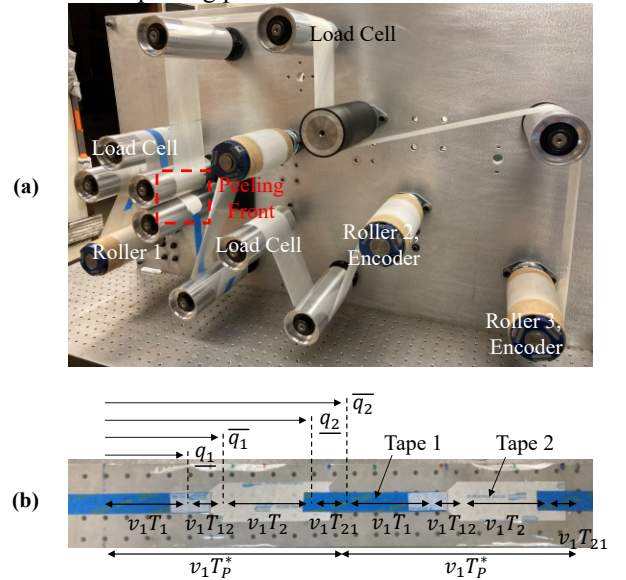

**Fig. 8.** R2R peeling experimental testbed. (a) The R2R machine; (b) The peeling pattern.

TABLE II
PARAMETERS OF THE EXPERIMENTAL SYSTEM

| $E_1, E_2, E_3$ (GPa) | $R_1, R_2, R_3$ (m) | $A_{r_1}, A_{r_2}, A_{r_3}$ (mm²) | $J_2, J_3$ (kg-m²) | $f_2, f_3$ (N-m-s-rad⁻¹) |
|---|---|---|---|---|
| 2.0 | 0.0381 | 15.5, 12.9, 12.9 | 0.65, 0.75 | 9.2, 11.4 |

TABLE III
THE THREE EXPERIMENTAL CASES

| Case | Setpoint | | | Peeling Pattern Properties | | |
|---|---|---|---|---|---|---|
| | $\theta$ (deg.) | $\alpha$ (deg.) | $v_1$ (cm/s) | Section | $\overline{q_i}, \underline{q_i}$ (cm) | $\overline{t_i}, \underline{t_i}$ (s) |
| 1 | 50 | 155 | 1.1 | Tape 1 | 16.5, 22.0 | 15, 20 |
| | | | | Tape 2 | 38.5, 44.0 | 35, 40 |
| 2 | 56 | 152 | .73 | Tape 1 | 11.0, 14.7 | 15, 20 |
| | | | | Tape 2 | 25.7, 29.3 | 35, 40 |
| 3 | 56 | 147 | .88 | Tape 1 | 13.2, 17.6 | 15, 20 |
| | | | | Tape 2 | 30.8, 35.2 | 35, 40 |

*E. Results and discussion*

Figure 9 shows one representative example trajectory for each of the three experimental cases. Figure 10 gives the measured $\tau$ signals for all the experimental runs. In the figure, the black dashed lines give the mode-dependent upper and lower bounds of $\tau$ that were used to generate the uncertain APPLS system model for the proposed controller. The heavy solid black lines are the uniform upper and lower bounds of $\tau$ that were used to generate the conservative uncertain system model for the baseline controller. Figure 11 gives box plots of the RMS errors of the two peeling angles and the standard deviations of the velocity and control signals for each case.

The example trajectory in Fig. 9 shows that the baseline controller induces far more undesired variations than the proposed controller. Furthermore, Fig. 11 shows that the proposed controller achieves less mean RMS angle error than the baseline controller, and the baseline controller induces significantly more velocity and control variation. Specifically, over the three cases, the mean RMS $\theta$ error is decreased by an average of 17.5% relative to the baseline, while the mean RMS $\alpha$ error is decreased by an average of 19.6%. This difference would have a significant impact on peeling performance of advanced materials [10], [13]. In addition, over the three cases, the mean $v_2$, $v_3$, $u_2$, and $u_3$ standard deviations are decreased by an average of 69.9%, 67.5%, 63.5%, and 62.7%, respectively, relative to the baseline. Over time, excessive variations in web velocity and motor torque would result in significant wear to process components and actuators. Thus, the proposed controller achieves significantly better performance than the baseline controller.

The baseline controller results in worse performance because it is highly conservative. Specifically, because the controller synthesis procedure does not account for the peeling pattern, the controller gain must be made too high to guarantee the same $\mathcal{H}_\infty$ performance the proposed controller. This unnecessarily high-gain feedback excites unmodeled dynamics, resulting in severe oscillations. Additionally, as shown in Section IV.B, the dwell-time and modeling uncertainty included in the uncertain APPLS structure is necessary to account for the nonlinearities and parameter variance inherent in the system. The results show that the proposed uncertain APPLS $\mathcal{H}_\infty$ analysis and control design procedure finds the optimal tradeoff by considering the correct amount of uncertainty to ensure that the necessary performance is achieved, without being too conservative such that a high control gain would harm system performance.

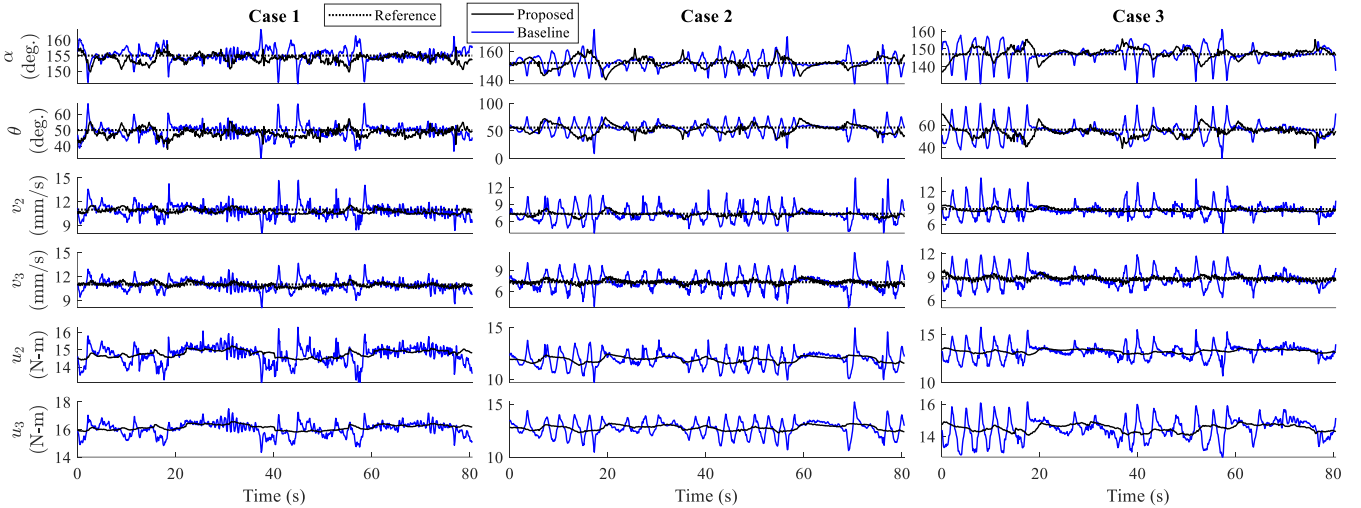

**Fig 9.** Example experimental state and control trajectories

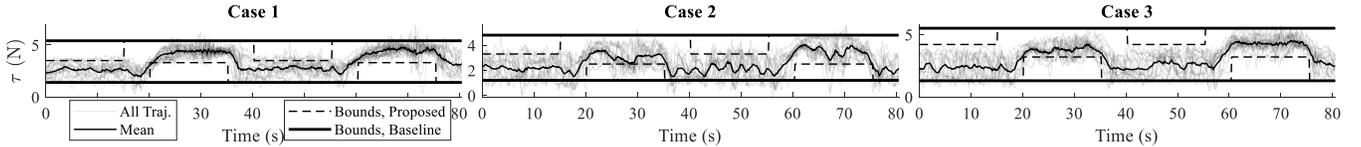

**Fig. 10.** Experimental $\tau$ variation

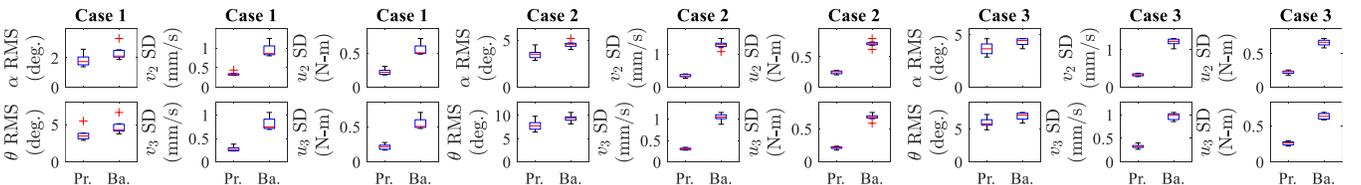

**Fig. 11.** Angle errors, velocity variation, and control variation. "Pr." stands for the proposed controller and "Ba." stands for the baseline controller.

## V. CONCLUSION

This study presents a novel switched system analysis and control framework for uncertain APPLSs with diverse applications. The APPLS structure extends the standard PPLS by incorporating dwell-time uncertainty, while the uncertain APPLS is a further generalization that accounts for modeling uncertainty. To facilitate controller synthesis, an $\mathcal{H}_\infty$ performance result is established for APPLSs and uncertain APPLSs. The theoretical treatment utilizes the system switching structure and a discontinuous, time-varying, MDLF. The novel $\mathcal{H}_\infty$ performance result is leveraged to develop an $\mathcal{H}_\infty$ controller synthesis algorithm for uncertain APPLSs. The resulting control policy can achieve optimal performance without costly online computation. This novel $\mathcal{H}_\infty$ controller is applied to R2R dry transfer of patterned 2D materials. Experiments were conducted and the results demonstrate that the proposed method enables a less conservative and, therefore, better-performing controller. The proposed controller reduced the average peeling angle error by 19% and the average control input and motor speed variation by 69% and 63%, respectively. This improvement in performance will significantly increase the transfer quality and drastically reduce component wear.

Future work could relax the current modeling assumptions for each mode, thus extending the almost periodic piecewise switching framework to a broader class of systems. For example, the assumption of known nominal dynamics could be lifted, enabling the use of adaptive or model-free methods that learn the dynamics or optimal control policy for each mode online. Additionally, alternative uncertainty structures—such as structured, multiplicative, or time-varying uncertainty—could be introduced to match a given application. Likewise, with advancements in machine learning, data-driven techniques could be integrated to enhance control of processes with an APPLS structure, such as R2R dry transfer. Hybridizing such a data-driven approach with a-priori knowledge of system dynamics could further improve accuracy while maintaining understanding of the process physics. The work presented in this study provides critical foundation for these potential extensions.

## APPENDIX

**Proof of Theorem 1**: First, Eqs. (4)-(6) guarantee that the uncertain APPLS in Eq. (1), controlled using the control structure in Eq. (2), is $\lambda^*$-exponentially stable when $w = 0$ [6], [7]. This stability result can be proven using the MDLF in Eq. (7) [6], [7].

Next, Eq. (4) can be rearranged into the following set of LMIs.

$$\begin{bmatrix} \mathbf{Z}_i & * & * \\ (\mathbf{P}_i \mathbf{B}_{w_i})^T & -\gamma^2 \mathbf{I} & * \\ (\mathbf{P}_i \mathbf{F}_i)^T & 0 & -\beta_i \mathbf{I} \end{bmatrix} \leq 0 \quad (21.A)$$

$$\begin{bmatrix} \mathbf{Z}_{i,i+1} & * & * \\ (\mathbf{P}_{i,i+1} \mathbf{B}_{w_i})^T & -\gamma^2 \mathbf{I} & * \\ (\mathbf{P}_{i,i+1} \mathbf{F}_i)^T & 0 & -\beta_{i,i+1} \mathbf{I} \end{bmatrix} \leq 0 \quad (21.B)$$

$$\begin{bmatrix} \mathbf{Z}_{i+1,i+1} & * & * \\ (\mathbf{P}_{i,i+1} \mathbf{B}_{w_{i+1}})^T & -\gamma^2 \mathbf{I} & * \\ (\mathbf{P}_{i,i+1} \mathbf{F}_{i+1})^T & 0 & -\beta_{i+1,i+1} \mathbf{I} \end{bmatrix} \leq 0 \quad (21.C)$$

where $\mathbf{Z}_i = sym(\mathbf{P}_i(\mathbf{A}_i + \mathbf{B}_i \mathbf{K}_i)) + \lambda_i \mathbf{P}_i + (\mathbf{C}_i + \mathbf{D}_i \mathbf{K}_i)^T (\mathbf{C}_i + \mathbf{D}_i \mathbf{K}_i) + \beta_i \mathbf{G}_i^T \mathbf{G}_i$, $\mathbf{Z}_{i,i+1} = sym(\mathbf{P}_{i,i+1}(\mathbf{A}_i + \mathbf{B}_i \mathbf{K}_{i,i+1})) + \lambda_{i,i+1} \mathbf{P}_{i,i+1} + (\mathbf{C}_i + \mathbf{D}_i \mathbf{K}_{i,i+1})^T (\mathbf{C}_i + \mathbf{D}_i \mathbf{K}_{i,i+1}) + \beta_{i,i+1} \mathbf{G}_i^T \mathbf{G}_i$, $\mathbf{Z}_{i+1,i+1} = sym(\mathbf{P}_{i,i+1}(\mathbf{A}_{i+1} + \mathbf{B}_{i+1} \mathbf{K}_{i,i+1})) + \lambda_{i,i+1} \mathbf{P}_{i,i+1} + (\mathbf{C}_{i+1} + \mathbf{D}_{i+1} \mathbf{K}_{i,i+1})^T (\mathbf{C}_{i+1} + \mathbf{D}_{i+1} \mathbf{K}_{i,i+1}) + \beta_{i+1,i+1} \mathbf{G}_{i+1}^T \mathbf{G}_{i+1}$, $\mathbf{P}_i = \mathbf{Q}_i^{-1}$, $\mathbf{P}_{i,i+1} = \mathbf{Q}_{i,i+1}^{-1}$, $\mathbf{K}_i = \mathbf{Y}_i \mathbf{Q}_i^{-1}$, $\mathbf{K}_{i,i+1} = \mathbf{Y}_{i,i+1} \mathbf{Q}_{i,i+1}^{-1}$, $\beta_i = \alpha_i^{-1}$, $\beta_{i,i+1} = \alpha_{i,i+1}^{-1}$, and $\beta_{i,i+1} = \alpha_{i,i+1}^{-1}$. Using the $\mathcal{S}$-procedure [33], Eq. (21) implies that, for $t \in [lT_P^* + \overline{t_{i-1}}, lT_P^* + \underline{t_i})$,

$$\overline{\mathbf{x}^T \mathbf{P}_i \mathbf{x}} + \lambda_i \mathbf{x}^T \mathbf{P}_i \mathbf{x} + \mathbf{z}^T \mathbf{z} - \gamma^2 \mathbf{w}^T \mathbf{w} \leq 0 \quad (22)$$

and for $t \in [lT_P^* + \underline{t_i}, lT_P^* + t_{l,i}) \cup [lT_P^* + t_{l,i}, lT_P^* + \overline{t_i})$,

$$\overline{\mathbf{x}^T \mathbf{P}_{i,i+1} \mathbf{x}} + \lambda_{i,i+1} \mathbf{x}^T \mathbf{P}_{i,i+1} \mathbf{x} + \mathbf{z}^T \mathbf{z} - \gamma^2 \mathbf{w}^T \mathbf{w} \leq 0 \quad (23)$$

Then, Eqs. (22) and (23) lead directly to the following bounds on the MDLF in Eq. (7). For $t \in [lT_P^* + \overline{t_{i-1}}, lT_P^* + \underline{t_i})$,

$$\dot{V}(t) \leq -\lambda_i V(t) - (\mathbf{z}^T \mathbf{z} - \gamma^2 \mathbf{w}^T \mathbf{w}) \quad (24)$$

and for $t \in [lT_P^* + \underline{t_i}, lT_P^* + \overline{t_i})$,

$$\dot{V}(t) \leq -\lambda_{i,i+1} V(t) - (\mathbf{z}^T \mathbf{z} - \gamma^2 \mathbf{w}^T \mathbf{w}) \quad (25)$$

Next, Eq. (5) can be rearranged to yield the following relationships.

$$\mathbf{P}_1 \leq \mu_1 \mathbf{P}_{S,S+1} \quad (26.A)$$
$$\mathbf{P}_i \leq \mu_i \mathbf{P}_{i-1,i} \quad (26.B)$$
$$\mathbf{P}_{i,i+1} \leq \mu_{i,i+1} \mathbf{P}_i \quad (26.C)$$

where $\mathbf{P}_i = \mathbf{Q}_i^{-1}$ and $\mathbf{P}_{i,i+1} = \mathbf{Q}_{i,i+1}^{-1}$. Then, using Eqs. (24)-(26), and the definition of $V(t)$ in Eq. (7), the following bound on $V(t)$ for $t \in [lT_P^* + \underline{t_m}, lT_P^* + \overline{t_m})$ can be established, relative to the initial condition $V(0)$, using convolution.

$$V(t) \leq e^{\Theta_0} V(0) - \sum_{j=1}^{l} \sum_{k=1}^{S} \left[ \int_{(j-1)T_P^* + \overline{t_{k-1}}}^{(j-1)T_P^* + \underline{t_k}} e^{\Theta_1} \mathcal{F}(\tau) d\tau \right.$$
$$\left. + \int_{(j-1)T_P^* + \underline{t_k}}^{(j-1)T_P^* + \overline{t_k}} e^{\Theta_2} \mathcal{F}(\tau) d\tau \right]$$
$$- \sum_{k=1}^{m-1} \left[ \int_{lT_P^* + \overline{t_{k-1}}}^{lT_P^* + \underline{t_k}} e^{\Theta_3} \mathcal{F}(\tau) d\tau + \int_{lT_P^* + \underline{t_k}}^{lT_P^* + \overline{t_k}} e^{\Theta_4} \mathcal{F}(\tau) d\tau \right] \quad (27)$$
$$- \int_{lT_P^* + \overline{t_{m-1}}}^{lT_P^* + \underline{t_m}} e^{\Theta_5} \mathcal{F}(\tau) d\tau - \int_{lT_P^* + \underline{t_m}}^{t} e^{\Theta_6} \mathcal{F}(\tau) d\tau$$

where $\mathcal{F}(t) = \mathbf{z}^T(t)\mathbf{z}(t) - \gamma^2 \mathbf{w}^T(t)\mathbf{w}(t)$, and the $\Theta_i$, $i = 0, 1, \ldots, 6$ terms are defined below.

$$\Theta_0 = \sum_{i=1}^{m} M_i + (l-1)\ln\mu_1 + l\ln\mu_{1,2} + l\sum_{i=2}^{S} M_i - \qquad (28.A)$$
$$l\sum_{i=1}^{S} \Lambda_i - \sum_{i=1}^{m-1} \Lambda_i - \lambda_m T_m - \lambda_{m,m+1}\left(t - \left(lT_P^* + \underline{t_m}\right)\right)$$

$$\Theta_1 = \sum_{i=1}^{m} M_i + (l-j+1)\sum_{i=k+1}^{S} M_i + (l-j)\sum_{i=1}^{k} M_i$$
$$+ \ln\mu_{k,k+1} - \lambda_k\left((j-1)T_P^* + \underline{t_k} - \tau\right) - \qquad (28.B)$$
$$\sum_{i=k+1}^{S} \lambda_i T_i + \lambda_{i-1,i}T_{i-1,i} - \lambda_{S,S+1}T_{S,S+1} - (l-j)\sum_{i=1}^{S} \Lambda_i$$
$$- \sum_{i=1}^{m-1} \Lambda_i - \lambda_m T_m - \lambda_{m,m+1}\left(t - \left(lT_P^* + \underline{t_m}\right)\right)$$

$$\Theta_2 = \sum_{i=1}^{m} M_i + (l-j+1)\sum_{i=k+1}^{S} M_i + (l-j)\sum_{i=1}^{k} M_i -$$
$$\lambda_{k,k+1}\left((j-1)T_P^* + \overline{t_k} - \tau\right) - \sum_{i=k+1}^{S} \Lambda_i - (l-j)\sum_{i=1}^{S} \Lambda_i \qquad (28.C)$$
$$- \sum_{i=1}^{m-1} \Lambda_i - \lambda_m T_m - \lambda_{m,m+1}\left(t - \left(lT_P^* + \underline{t_m}\right)\right)$$

$$\Theta_3 = \sum_{i=k+1}^{m} M_i + \ln\mu_{k,k+1} - \lambda_k\left(lT_P^* + \underline{t_k} - \tau\right) - \qquad (28.D)$$
$$\sum_{i=k+1}^{m} \lambda_i T_i + \lambda_{i-1,i}T_{i-1,i} - \lambda_{m,m+1}\left(t - \left(lT_P^* + \underline{t_m}\right)\right)$$

$$\Theta_4 = \sum_{i=k+1}^{m} M_i - \lambda_{k,k+1}\left(lT_P^* + \overline{t_k} - \tau\right) - \sum_{i=k+1}^{m-1} \Lambda_i \qquad (28.E)$$
$$- \lambda_m T_m - \lambda_{m,m+1}\left(t - \left(lT_P^* + \underline{t_m}\right)\right)$$

$$\Theta_5 = \ln\mu_{m,m+1} - \lambda_m\left(lT_P^* + \underline{t_m} - \tau\right) \qquad (28.F)$$
$$- \lambda_{m,m+1}\left(t - \left(lT_P^* + \underline{t_m}\right)\right)$$

$$\Theta_6 = -\lambda_{m,m+1}(t-\tau) \qquad (28.G)$$

In this way, the MDLF in Eq. (7) is bounded by a stable linear system that is solved explicitly, creating a homogenous term and a particular term that depends on the difference between the weighted input and output norms, $\mathcal{F}(t)$.

Next, to get an inequality that bounds the weighted signal norm of $\mathbf{z}$ by the signal norm of $\mathbf{w}$, note that $V(t) > 0$, and multiply both sides of Eq. (27) by $e^{\Upsilon}$, where $\Upsilon = -(l-1)\ln\mu_1 - l\ln\mu_{1,2} - l\sum_{i=2}^{S} M_i - \sum_{i=1}^{m} M_i = -l\sum_{i=1}^{S} M_i - \sum_{i=2}^{m} M_i - \ln\mu_{1,2}$. This gives the following relationship.

$$\sum_{j=1}^{l}\sum_{k=1}^{S}\left[\int_{(j-1)T_P^*+\overline{t_{k-1}}}^{(j-1)T_P^*+\underline{t_k}} e^{\Psi_1}\|z(\tau)\|^2 d\tau + \right.$$
$$\int_{(j-1)T_P^*+\underline{t_k}}^{(j-1)T_P^*+\overline{t_k}} e^{\Psi_2}\|z(\tau)\|^2 d\tau\right] + \sum_{k=1}^{m-1}\left[\int_{lT_P^*+\overline{t_{k-1}}}^{lT_P^*+\underline{t_k}} e^{\Psi_3}\|z(\tau)\|^2 d\tau\right.$$
$$+ \int_{lT_P^*+\underline{t_k}}^{lT_P^*+\overline{t_k}} e^{\Psi_4}\|z(\tau)\|^2 d\tau\right] + \int_{lT_P^*+\overline{t_{m-1}}}^{lT_P^*+\underline{t_m}} e^{\Psi_5}\|z(\tau)\|^2 d\tau$$
$$+ \int_{lT_P^*+\underline{t_m}}^{t} e^{\Psi_6}\|z(\tau)\|^2 d\tau \leq e^{\Xi_0}V(0) \qquad (29)$$
$$+ \gamma^2\left\{\sum_{j=1}^{l}\sum_{k=1}^{S}\left[\int_{(j-1)T_P^*+\overline{t_{k-1}}}^{(j-1)T_P^*+\underline{t_k}} e^{\Xi_1}\|w(\tau)\|^2 d\tau + \right.\right.$$
$$\int_{(j-1)T_P^*+\underline{t_k}}^{(j-1)T_P^*+\overline{t_k}} e^{\Xi_2}\|w(\tau)\|^2 d\tau\right] + \sum_{k=1}^{m-1}\left[\int_{lT_P^*+\overline{t_{k-1}}}^{lT_P^*+\underline{t_k}} e^{\Xi_3}\|w(\tau)\|^2 d\tau\right.$$
$$+ \int_{lT_P^*+\underline{t_k}}^{lT_P^*+\overline{t_k}} e^{\Xi_4}\|w(\tau)\|^2 d\tau\right] + \int_{lT_P^*+\overline{t_{m-1}}}^{lT_P^*+\underline{t_m}} e^{\Xi_5}\|w(\tau)\|^2 d\tau$$
$$\left.+ \int_{lT_P^*+\underline{t_m}}^{t} e^{\Xi_6}\|w(\tau)\|^2 d\tau\right\}$$

where the $\Xi_i$, $i = 0,1,\ldots,6$ and $\Psi_i$, $i = 1,2,\ldots,6$ terms are defined below.

$$\Xi_0 = -l\sum_{i=1}^{S} \Lambda_i - \sum_{i=1}^{m-1} \Lambda_i - \lambda_m T_m \qquad (30.A)$$
$$- \lambda_{m,m+1}\left(t - \left(lT_P^* + \underline{t_m}\right)\right)$$

$$\Xi_1 = -\sum_{i=k+1}^{S} \lambda_i T_i + \lambda_{i-1,i}T_{i-1,i} - \lambda_{S,S+1}T_{S,S+1}$$
$$- \lambda_k\left((j-1)T_P^* + \underline{t_k} - \tau\right) - (l-j)\sum_{i=1}^{S} \Lambda_i - \sum_{i=1}^{m-1} \Lambda_i \qquad (30.B)$$
$$- \lambda_m T_m - \lambda_{m,m+1}\left(t - \left(lT_P^* + \underline{t_m}\right)\right)$$

$$\Xi_2 = -\lambda_{k,k+1}\left((j-1)T_P^* + \overline{t_k} - \tau\right) - \sum_{i=k+1}^{S} \Lambda_i$$
$$- (l-j)\sum_{i=1}^{S} \Lambda_i - \sum_{i=1}^{m-1} \Lambda_i \qquad (30.C)$$
$$- \lambda_m T_m - \lambda_{m,m+1}\left(t - \left(lT_P^* + \underline{t_m}\right)\right)$$

$$\Xi_3 = -\lambda_k\left(lT_P^* + \underline{t_k} - \tau\right) - \sum_{i=k+1}^{m} \lambda_i T_i + \lambda_{i-1,i}T_{i-1,i} \qquad (30.D)$$
$$- \lambda_{m,m+1}\left(t - \left(lT_P^* + \underline{t_m}\right)\right)$$

$$\Xi_4 = -\lambda_{k,k+1}\left(lT_P^* + \overline{t_k} - \tau\right) - \sum_{i=k+1}^{m-1} \Lambda_i \qquad (30.E)$$
$$- \lambda_m T_m - \lambda_{m,m+1}\left(t - \left(lT_P^* + \underline{t_m}\right)\right)$$

$$\Xi_5 = -\lambda_m\left(lT_P^* + \underline{t_m} - \tau\right) \qquad (30.F)$$
$$- \lambda_{m,m+1}\left(t - \left(lT_P^* + \underline{t_m}\right)\right)$$

$$\Xi_6 = -\lambda_{m,m+1}(t-\tau) \qquad (30.G)$$

$$\Psi_1 = \Xi_1 - (j-1)\sum_{i=1}^{S} M_i - \ln\mu_{1,2} + \sum_{i=2}^{k-1} M_i - \ln\mu_k \quad (31.A)$$

$$\Psi_2 = \Xi_2 - (j-1)\sum_{i=1}^{S} M_i - \sum_{i=2}^{k} M_i - \ln\mu_{1,2} \quad (31.B)$$

$$\Psi_3 = \Xi_3 - l\sum_{i=1}^{S} M_i - \ln\mu_{1,2} - \sum_{i=2}^{k-1} M_i - \ln\mu_k \quad (31.C)$$

$$\Psi_4 = \Xi_4 - l\sum_{i=1}^{S} M_i - \ln\mu_{1,2} - \sum_{i=2}^{k} M_i \quad (31.D)$$

$$\Psi_5 = \Xi_5 - l\sum_{i=1}^{S} M_i - \ln\mu_{1,2} - \sum_{i=2}^{m-1} M_i - \ln\mu_m \quad (31.E)$$

$$\Psi_6 = \Xi_6 - l\sum_{i=1}^{S} M_i - \ln\mu_{1,2} - \sum_{i=2}^{m} M_i \quad (31.F)$$

To combine terms in Eq. (29), starting with the definitions of $\Psi_i, i = 0, 1, \ldots, 6$ and $\Xi_i, i = 1, 2, \ldots, 6$, the following chain of inequalities can be proven.

$$\begin{aligned}
\Psi_1 &\geq -\lambda_{max}\Big[(j-1)T_P^* + \underline{t_k} - \tau + \sum_{i=k+1}^{S} T_i + T_{i-1,i} \\
&+ T_{S,S+1} + (l-j)\sum_{i=1}^{S} T_i + T_{i,i+1} + \sum_{i=1}^{m-1} T_i + T_{i,i+1} \\
&+ T_m + t - lT_P^* - \underline{t_m}\Big] - j\sum_{i=1}^{S} M_i \\
&\geq -\lambda_{max}[T_P^* - jT_P^* + t] + j(-\lambda_{max}T_P^* + 2\lambda^* T_P^*) \\
&\geq -\lambda_{max}(T_P^* + t) + 2\lambda^* T_P^*
\end{aligned} \quad (32)$$

By similar logic, it can be shown that $\Psi_2, \Psi_3, \Psi_4, \Psi_5,$ and $\Psi_6 \geq -\lambda_{max}(T_P^* + t) + 2\lambda^* T_P^*$. Additionally,

$$\begin{aligned}
\Xi_0 &\leq -l2\lambda^* T_P^* \\
&-\lambda_{min}\Big[\sum_{i=1}^{m-1} T_i + T_{i,i+1} + T_m + t - lT_P^* - \underline{t_m}\Big] \\
&= -2\lambda^* t + (t - lT_P^*)(2\lambda^* - \lambda_{min}) \\
&\leq -2\lambda^* t + (2\lambda^* - \lambda_{min})T_P^*
\end{aligned} \quad (33)$$

$$\begin{aligned}
\Xi_2 &\leq -\lambda_{min}\Big[(j-1)T_P^* + \overline{t_k} - \tau + \sum_{i=k+1}^{S} T_i + T_{i,i+1} + \\
&\sum_{i=1}^{m-1} T_i + T_{i,i+1} + T_m + t - lT_P^* - \underline{t_m}\Big] - (l-j)\sum_{i=1}^{S} \Lambda_i \\
&\leq -2\lambda^*(t-\tau) + (2\lambda^* - \lambda_{min})\big((t-\tau) - (l-j)T_P^*\big) \\
&\leq -2\lambda^*(t-\tau) + (2\lambda^* - \lambda_{min})2T_P^*
\end{aligned} \quad (34)$$

By similar logic, it can be shown that $\Xi_1, \Xi_3, \Xi_4, \Xi_5,$ and $\Xi_6 \leq -2\lambda^*(t-\tau) + (2\lambda^* - \lambda_{min})2T_P^*$. Finally, these bounds on the exponential coefficients allow Eq. (29) to be simplified as follows,

$$\begin{aligned}
\int_0^t e^{(-\lambda_{max}(t+T_P^*) + 2\lambda^* T_P^*)} \mathbf{z}^T(\tau)\mathbf{z}(\tau)d\tau &\leq \\
e^{(-2\lambda^* t + (2\lambda^* - \lambda_{min})T_P^*)}V(0) &+ \\
\gamma^2 \int_0^t e^{(-2\lambda^*(t-\tau) + (2\lambda^* - \lambda_{min})2T_P^*)} \mathbf{w}^T(\tau)\mathbf{w}(\tau)d\tau &
\end{aligned} \quad (35)$$

Then, integrating $t$ from 0 to $\infty$ and dividing by $e^{2\lambda^* T_P^*}$, the following relationship follows.

$$\int_0^\infty e^{-\lambda\tau} \mathbf{z}^T(\tau)\mathbf{z}(\tau)d\tau \leq aV(0) + b\gamma^2 \int_0^\infty \mathbf{w}^T(\tau)\mathbf{w}(\tau)d\tau \quad (36)$$

where $a = \frac{\lambda_{max}}{2\lambda^*}e^{((\lambda_{max}-\lambda_{min})T_P^*)}$, and $b$ and $\lambda$ are defined in the theorem statement. Equation (36) is a weighted $\mathcal{L}_2$-gain criterion for the system [16], [17]. Assuming zero initial conditions, Eq. (36) is equivalent to the weighted $\mathcal{H}_\infty$ performance result in Eq. (3). □

## References

[1] J. Jiao et al., "Dynamics of a periodic switched predator–prey system with impulsive harvesting and hibernation of prey population," *J Franklin Inst*, vol. 353, no. 15, pp. 3818–3834, Oct. 2016.

[2] C. W. Wong et al., "Periodic forced vibration of unsymmetrical piecewise-linear systems by incremental harmonic balance method," *J Sound Vib*, vol. 149, no. 1, pp. 91–105, Aug. 1991.

[3] Jian Sun et al., "Averaged modeling of PWM converters operating in discontinuous conduction mode," *IEEE Trans Power Electron*, vol. 16, no. 4, pp. 482–492, Jul. 2001.

[4] S. Teng et al., "A Least-Laxity-First Scheduling Algorithm of Variable Time Slice for Periodic Tasks," *International Journal of Software Science and Computational Intelligence*, vol. 2, no. 2, pp. 86–104, Apr. 2010.

[5] B. Wei et al., "Stability Analysis of Continuous-Time Almost Periodic Piecewise Linear Systems with Dwell Time Uncertainties," in *Proceedings of the 43rd Chinese Control Conference*, 2024, pp. 1153–1159.

[6] C. Martin et al., "Stabilization of Almost Periodic Piecewise Linear Systems with Norm-Bounded Uncertainty for Roll-to-Roll Dry Transfer Manufacturing Processes," in *2024 American Control Conference (ACC)*, 2024, pp. 1121–1126.

[7] C. Fan et al., "Stability and Stabilization of Almost Periodic Piecewise Linear Systems With Dwell Time Uncertainty," *IEEE Trans Automat Contr*, vol. 68, no. 2, pp. 1130–1137, Feb. 2023.

[8] N. Hong et al., "The Line Speed Effect in Roll-to-Roll Dry Transfer of Chemical Vapor Deposition Graphene," *International Conference on Micro- and Nano-devices Enabled by R2R Manufacturing*, 2021.

[9] N. Hong et al., "A method to estimate adhesion energy of as-grown graphene in a roll-to-roll dry transfer process," *Carbon N Y*, vol. 201, pp. 712–718, Jan. 2023.

[10] N. Hong et al., "Roll-to-Roll Dry Transfer of Large-Scale Graphene," *Advanced Materials*, vol. 34, no. 3, Jan. 2022.

[11] S. R. Na et al., "Selective mechanical transfer of graphene from seed copper foil using rate effects," *ACS Nano*, vol. 9, no. 2, pp. 1325–1335, Feb. 2015.

[12] H. Xin et al., "Roll-to-roll mechanical peeling for dry transfer of chemical vapor deposition graphene," *J Micro Nanomanuf*, vol. 6, no. 3, Sep. 2018.

[13] J. Zhao et al., "Patterned Peeling 2D MoS2 off the Substrate," *ACS Appl Mater Interfaces*, vol. 8, no. 26, pp. 16546–16550, Jul. 2016.

[14] K. H. Choi et al., "Web register control algorithm for roll-to-roll system based printed electronics," in *2010 IEEE International Conference on Automation Science and Engineering*, 2010.

[15] C. Martin et al., "Optimal Control of a Roll-to-Roll Dry Transfer Process With Bounded Dynamics Convexification," *J Dyn Syst Meas Control*, vol. 147, no. 3, p. 031004, May 2025.

[16] P. Li et al., "Stability, stabilization and L2-gain analysis of periodic piecewise linear systems," *Automatica*, vol. 61, pp. 218–226, Nov. 2015.

[17] Panshuo Li et al., "Stability and L2-gain analysis of periodic piecewise linear systems," in *2015 American Control Conference (ACC)*, 2015, pp. 3509–3514.

[18] C. Martin et al., "H∞ Optimal Control for Maintaining the R2R Peeling Front," *IFAC-PapersOnLine*, vol. 55, no. 37, pp. 663–668, 2022.

[19] X. Xie et al., "Finite-time H∞ control of periodic piecewise linear systems," *Int J Syst Sci*, vol. 48, no. 11, pp. 2333–2344, Aug. 2017.

[20] P. Li et al., "H∞ control of periodic piecewise vibration systems with actuator saturation," *JVC/Journal of Vibration and Control*, vol. 23, no. 20, pp. 3377–3391, Dec. 2017.

[21] X. Xie et al., "H∞ control problem of linear periodic piecewise time-delay systems," *Int J Syst Sci*, vol. 49, no. 5, pp. 997–1011, Apr. 2018.

11[22] P. Li et al., "Stability and L2 Synthesis of a Class of Periodic Piecewise Time-Varying Systems," *IEEE Trans Automat Contr*, vol. 64, no. 8, pp. 3378–3384, Aug. 2019.
[23] X. Xie et al., "Robust time-weighted guaranteed cost control of uncertain periodic piecewise linear systems," *Inf Sci (N Y)*, vol. 460–461, pp. 238–253, Sep. 2018.
[24] C. Fan et al., "Peak-to-peak filtering for periodic piecewise linear polytopic systems," *Int J Syst Sci*, vol. 49, no. 9, pp. 1997–2011, Jul. 2018.
[25] Y. Liu et al., "Stabilization and L2- gain performance of periodic piecewise impulsive linear systems," *IEEE Access*, vol. 8, pp. 200146–200156, 2020.
[26] A. I. Zecevic and D. D. Siljak, "Stabilization of nonlinear systems with moving equilibria," *IEEE Trans Automat Contr*, vol. 48, no. 6, pp. 1036–1040, Jun. 2003.
[27] Guang-Hong Yang and Dan Ye, "Reliable H∞ Control of Linear Systems With Adaptive Mechanism," *IEEE Trans Automat Contr*, vol. 55, no. 1, pp. 242–247, Jan. 2010.
[28] H. Zhang et al., "Online Adaptive Policy Learning Algorithm for H∞ State Feedback Control of Unknown Affine Nonlinear Discrete-Time Systems," *IEEE Trans Cybern*, vol. 44, no. 12, pp. 2706–2718, Dec. 2014.
[29] J. Wang et al., "Composite Antidisturbance Control for Hidden Markov Jump Systems With Multi-Sensor Against Replay Attacks," *IEEE Trans Automat Contr*, vol. 69, no. 3, pp. 1760–1766, Mar. 2024.
[30] Q. Zhao et al., "A Dynamic System Model for Roll-to-Roll Dry Transfer of Two-Dimensional Materials and Printed Electronics," *J Dyn Syst Meas Control*, vol. 144, no. 7, Jul. 2022.
[31] H. Xin et al., "Roll-to-Roll Mechanical Peeling for Dry Transfer of Chemical Vapor Deposition Graphene," *J Micro Nanomanuf*, vol. 6, no. 3, Sep. 2018.
[32] A. E. Bryson, *Control of spacecraft and aircraft*, vol. 41. Princeton university press Princeton, 1993.
[33] S. P. Boyd, *Linear matrix inequalities in system and control theory*. Society for Industrial and Applied Mathematics, 1994.